\documentclass[twocolumn]{aastex631}
\usepackage{acronym}
\usepackage{amsmath}
\usepackage{acronym}
\usepackage[T1]{fontenc}
\usepackage[utf8]{inputenc}
\usepackage{xcolor}
\usepackage{graphicx}
\usepackage{graphics}
\usepackage{amsfonts}
\usepackage{savesym}
\savesymbol{tablenum}
\usepackage{siunitx}
\restoresymbol{SIX}{tablenum}
\usepackage{tikz}
\usepackage{pgfplots}
\usepackage{ulem} % \sout{} (remove for final version)
\usepgfplotslibrary{groupplots}
\newlength\figureheight
\newlength\figurewidth

\newlength\figureheightmed
\newlength\figurewidthmed

\newlength\figureheightlarge
\newlength\figurewidthlarge

\begin{document}

\title{Revealing Phase Transition in Dense Matter with\\ Gravitational
Wave Spectroscopy of Binary Neutron Star Mergers}

\author{Pedro L.~\surname{Espino}}
%%\email{email address: pespino@berkeley.edu}
%%\thanks{N3AS fellow}
\affiliation{Institute for Gravitation \& the Cosmos, The
Pennsylvania State University, University Park, PA 16802} 
\affiliation{Department of Physics, University of California, Berkeley, 
CA 94720, USA}

\author{Aviral \surname{Prakash}}
\affiliation{Institute for Gravitation \& the Cosmos, The
Pennsylvania State University, University Park, PA 16802} 
\affiliation{Department of Physics, The Pennsylvania State
University, University Park, PA 16802}

\author{David \surname{Radice}}
\thanks{Alfred P.~Sloan fellow}
\affiliation{Institute for Gravitation \& the Cosmos, The
Pennsylvania State University, University Park, PA 16802} 
\affiliation{Department of Physics, The Pennsylvania State
University, University Park, PA 16802}
\affiliation{Department of Astronomy \& Astrophysics, The
Pennsylvania State University,University Park, PA 16802}

\author{Domenico \surname{Logoteta}}
\affiliation{Dipartimento di Fisica, Universit\`{a} di Pisa, Largo B.  Pontecorvo, 3 I-56127 Pisa, Italy}
\affiliation{INFN, Sezione di Pisa, Largo B. Pontecorvo, 3 I-56127 Pisa, Italy}

\begin{abstract}
We use numerical relativity simulations of binary neutron star 
mergers to show that high density deconfinement phase transitions (PTs) to quark
matter can be 
probed using multimodal postmerger gravitational wave (GW) spectroscopy. 
Hadron-quark PTs suppress the one-armed spiral instability in the 
remnant. This is manifested in an anti-correlation 
between the energy carried in the $l=2, m=1$ GW mode and energy density gap
which separates the two phases. Consequently, a single measurement of the 
signal-to-noise ratios of the $l=2, m=1$ and $l=2, m=2$ GW modes could constrain the 
energy density gap of the PT.
\end{abstract}

\keywords{Neutron stars --- Equation of state --- Gravitational waves --- Hydrodynamics --- Instabilities}

\section{Introduction} \label{sec:intro}
Binary neutron star (BNS) mergers produce some of the most extreme conditions
in nature, compressing matter to several times the nuclear density and to
temperatures of tens of MeV \citep{Perego:2019adq}. More extreme
conditions are only found in the early Universe and in the interior of
black holes (BHs).
Multimessenger observations of binary neutron star (BNS) mergers can be used to
probe the properties of matter in these conditions, providing a unique
avenue to study the non-perturbative regime of QCD \citep{Shibata:2005xz,
Hinderer:2009ca, Damour:2012yf, Sekiguchi:2011mc, Hotokezaka:2011dh,
Bauswein:2013jpa, Radice:2016rys, LIGOScientific:2017fdd,
Margalit:2017dij, Bauswein:2017vtn, Radice:2017lry, Most:2018eaw, Most:2019onn,
Bauswein:2018bma, Coughlin:2018fis, De:2018uhw, LIGOScientific:2018hze,
LIGOScientific:2018cki, Radice:2018ozg, Dietrich:2020efo,
Breschi:2021tbm, Breschi:2021xrx, Kashyap:2021wzs, Perego:2021mkd,
Fujimoto:2022xhv, Prakash:2021wpz}.

Presently, there are large uncertainties in the fundamental physics of 
strongly-interacting matter at densities of a few times nuclear saturation~\citep{Capano:2019eae, Pang:2021jta, Annala:2021gom}.
It is not even clear what the relevant degrees of freedom are for the densities
and temperatures reached in the core of remnant massive neutron stars (RMNS) of BNS mergers. It is
possible that matter remains composed of nucleons, together with leptons
(electrons, positrons, and muons) and photons \citep{Perego:2019adq,
Loffredo:2022prq}. The appearance of more exotic baryons, such as hyperons, is
not excluded \citep{Sekiguchi:2011mc, Radice:2016rys, Logoteta:2021iuy}. 
It is also possible for a
transition to the deconfined quark-gluon plasma phase to take place in BNS
mergers \citep{Most:2018eaw, Most:2019onn, Bauswein:2018bma, Prakash:2021wpz}. The determination of the state
of matter formed in BNS mergers is one of the most pressing scientific
objectives of multimessenger astronomy \citep{Evans:2021gyd, Lovato:2022vgq}.

Previous work has shown that the presence of phase transitions to
deconfined quarks can be revealed by a shift of the postmerger
gravitational wave (GW) peak frequency $f_2$ from the value expected for
hadronic equations of state (EOSs) \citep{Bauswein:2018bma,
Weih:2019xvw, Blacker:2020nlq, Kedia:2022nns}.  However, 
such frequency shifts can be degenerate with deviations from
universal relations due to hadronic physics or other effects
\citep{Most:2018eaw, Weih:2019xvw, Liebling:2020dhf, Prakash:2021wpz,
Fujimoto:2022xhv, Tootle:2022pvd}. It has also been suggested that the
presence of a phase transition could be inferred from a measurement of
the threshold mass for prompt collapse of BNS systems
\citep{Bauswein:2020aag, Bauswein:2020xlt, Perego:2021mkd,
Kashyap:2021wzs}. In this \emph{Letter}, we use 8 state-of-the-art
numerical relativity simulations to show, for the first time, that the
presence and strength of a QCD phase transition could be unambiguously
determined through \emph{multimodal}  GW spectroscopy of RMNS. Such
measurements will be possible with the next-generation of  GW
experiments like Cosmic Explorer~\citep{Reitze:2019iox}, Einstein
Telescope~\citep{Punturo:2010zz}, and NEMO~\citep{Ackley:2020atn}. 

\section{Methods}\label{sec:methods}

We consider binaries in quasi-circular orbits and eccentric
encounters on nearly parabolic orbits. Although BNS mergers with highly eccentric 
orbits are expected to be significantly more rare than those with quasi-circular 
inspirals, these events may still have appreciable rates of as high as 50 
${\rm Gpc}^{-3}$~${\rm yr}^{-1}$~\citep{Lee:2009ca, Paschalidis:2015mla}; we 
include results from both types of mergers to consider as wide a variety of 
scenarios as possible. Initial data for the
quasi-circular binaries is created using the conformal thin sandwich
formalism \citep{York:1998hy} and assuming a helical Killing vector and
irrotational flows. The resulting elliptic equations are solved using the
pseudo-spectral code \texttt{LORENE} \citep{Gourgoulhon:2000nn,
Taniguchi:2001qv, Taniguchi:2001ww}. Initial data for the eccentric
encounters is constructed by superimposing two isolated, boosted,
neutron stars, following~\citet{Radice:2016dwd}. The initial separation of the
stellar barycenters for parabolic encounters is set to \SI{100}{km}, which is sufficiently large so that the
level of constraint violation in the initial data is comparable to that
of the quasi-circular binaries~\citep{Radice:2016dwd}.

We perform BNS merger simulations using the \texttt{WhiskyTHC} code
\citep{Radice:2012cu, Radice:2013hxh, Radice:2013xpa}. \texttt{WhiskyTHC} makes
use of the \texttt{CTGamma} spacetime solver \citep{Pollney:2009yz},
which is a part of the \texttt{Einstein Toolkit}
\citep{EinsteinToolkit:2022_05}. The adaptive mesh refinement driver
\texttt{Carpet} \citep{Schnetter:2003rb} is used to generate the
dynamical grid structure employed in the simulations. All simulations
considered in the present work have been performed using at least two grid resolutions. 
Although there
are quantitative differences in the GW waveforms computed at
different resolutions, the qualitative features discussed here are
robust across all simulations. Unless otherwise specified, we
discuss results from simulations using the fiducial grid resolution 
(with grid spacing $\Delta x \simeq \SI{184.6}{m}$ in the finest refinement level). 
The grid structure for the 
simulations is described in detail 
in~\citet{Radice:2018pdn} and~\citet{Radice:2016dwd} for the quasi-circular
and eccentric simulations, respectively.

\begin{figure*}[htb]
\centering
\includegraphics[width=0.49 \textwidth]{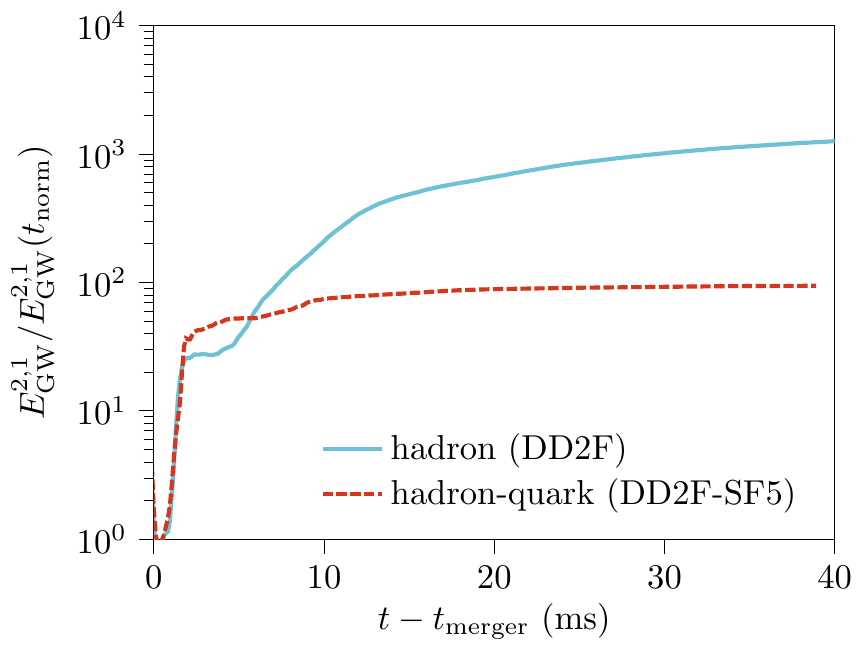}
\includegraphics[width=0.49 \textwidth]{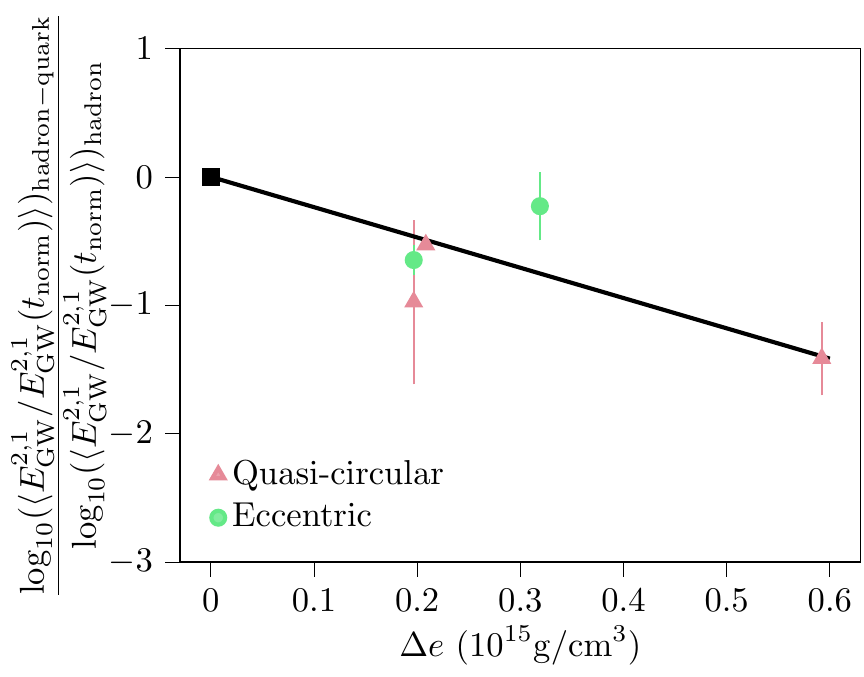}

\caption{{\it Left panel:} Energy 
carried by GWs in the $l=2,m=1$ mode as a function of time, scaled by the value at 
a fixed normalization time $t_{\rm norm} = t_{\rm mer} + \SI{0.5}{ms}$. The 
development of the one-armed spiral instability can be observed in the purely 
hadronic simulation as the energy in the 
$l=2,m=1$ GW mode continues to grow, but is suppressed in the hadron-quark 
simulation. 
{\it Right panel:} The same quantity depicted in the left panel, but 
time-averaged over a fixed time window and shown as a function of the energy 
density gap $\Delta e$ for each simulation.
We normalize by the same quantity for the complementary hadronic EOS. 
%%We depict results for 
%%quasi-circular and eccentric mergers with green circles and pink triangles, 
%%respectively. 
We include 
approximate error bars, obtained using lower resolution simulations, to represent 
the uncertainty introduced by our numerical methods. 
The black solid line represents a linear fit to the data from our simulations.
We find that the normalized energy emitted by the $l=2, m=1$ GW mode 
decreases for simulations that employ EOS models with larger values of $\Delta e$.
}
\label{fig:energy_GW_egap}
\end{figure*}

For a clear understanding of the role that high-density 
deconfinement phase transitions could play in the development of the 
one-armed spiral instability, we consider a total of 7 EOS models and run a total of 8 
simulations
with varying phase transition features. In particular, the 
size of the energy density gap which separates the hadronic and quark 
phases is a useful way to classify hybrid hadron-quark EOS models 
and provides a qualitative measure of the `strength' of the phase 
transition~\citep{Alford:2015gna}. 
As such, we 
consider EOS models that 
cover several sizes of the energy density gap, ranging from 
non-existent (i.e., a purely hadronic EOS) to large, while maintaining 
consistency with current astrophysical constraints on the dense matter 
EOS. We consider both phenomenological EOS 
models~\citep{Paschalidis:2017qmb, T91, T92, Bozzola:2019tit, 
Espino:2021adh} (in the form of 
piecewise polytropic approximations using the prescription of~\citet{Read2009}) and 
microphysical, 
finite temperature EOS models~\citep{PhysRevD.103.023001, Bombaci:2018ksa, 
Logoteta:2020yxf, Prakash:2021wpz}. We only consider equal-mass ratio 
binary configurations, with the total binary mass ranging from 
$\SI{2.6}{M_\odot} - \SI{2.7}{M_\odot}$. The lack of $\pi$-rotational
symmetry in BNS configurations 
with unequal-masses may be a suitable way of effectively seeding non-axisymmetric 
fluid instabilities that can take hold in the post-merger environment.
Neutrino emission and reabsorption are not
included for binaries in eccentric orbits, while all quasi-circular binaries include a 
neutrino treatment via the moment based M0 scheme \citep{Radice:2018pdn}. However, neutrinos
are not expected to influence the dynamics on the time scales considered in our study
\citep{Radice:2020ddv, Radice:2021jtw}. Additionally, magnetic fields are not 
accounted for in any of our simulations, but these are also expected to be subdominant \citep{Palenzuela:2021gdo}. We find that, despite the diversity in binary 
properties and differences in the evolution, the effects presented in this work are 
robust.

\section{Results} \label{sec:results}

\begin{figure*}[htb]
\centering
\includegraphics[width=0.49 \textwidth]{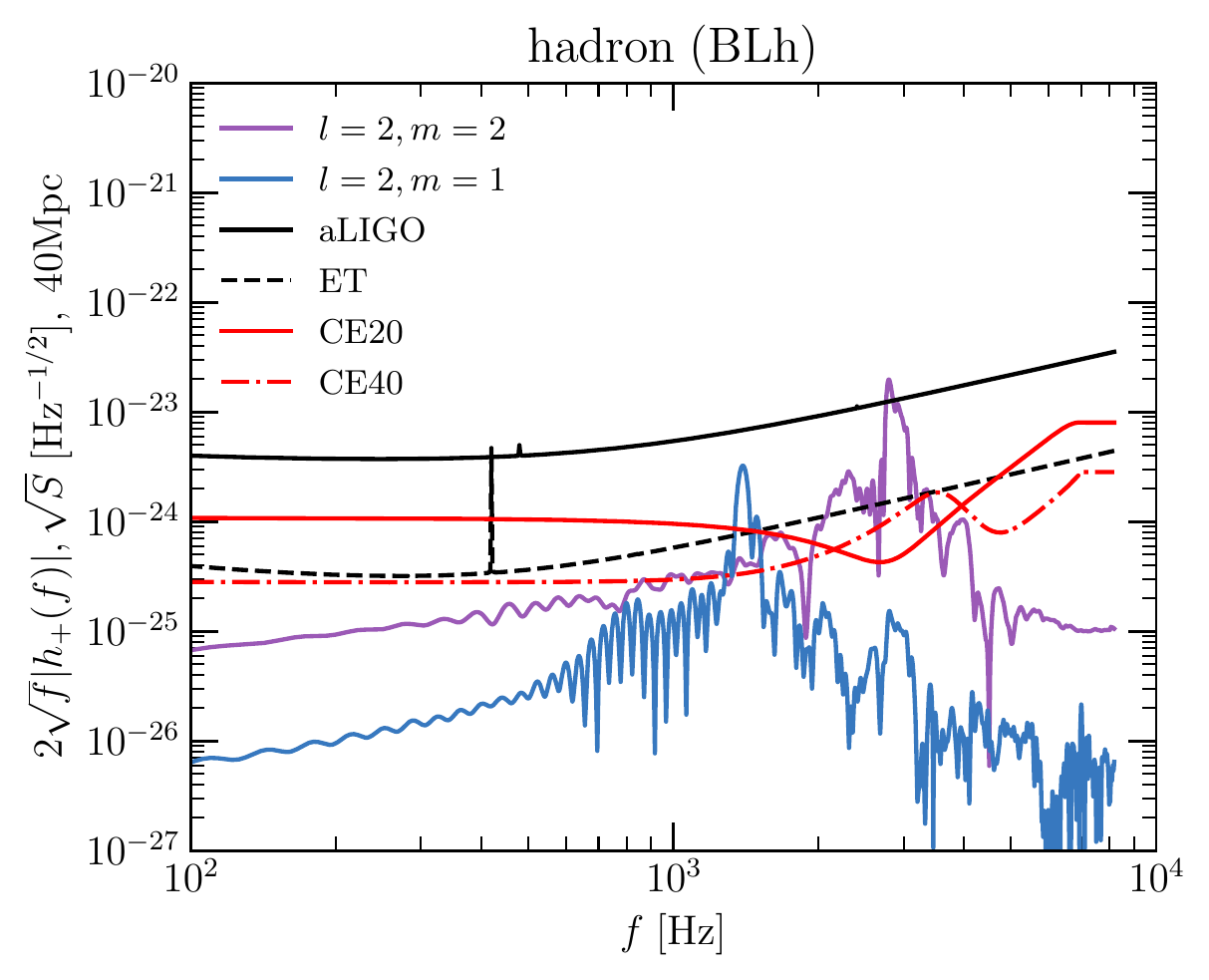}
\includegraphics[width=0.49 \textwidth]{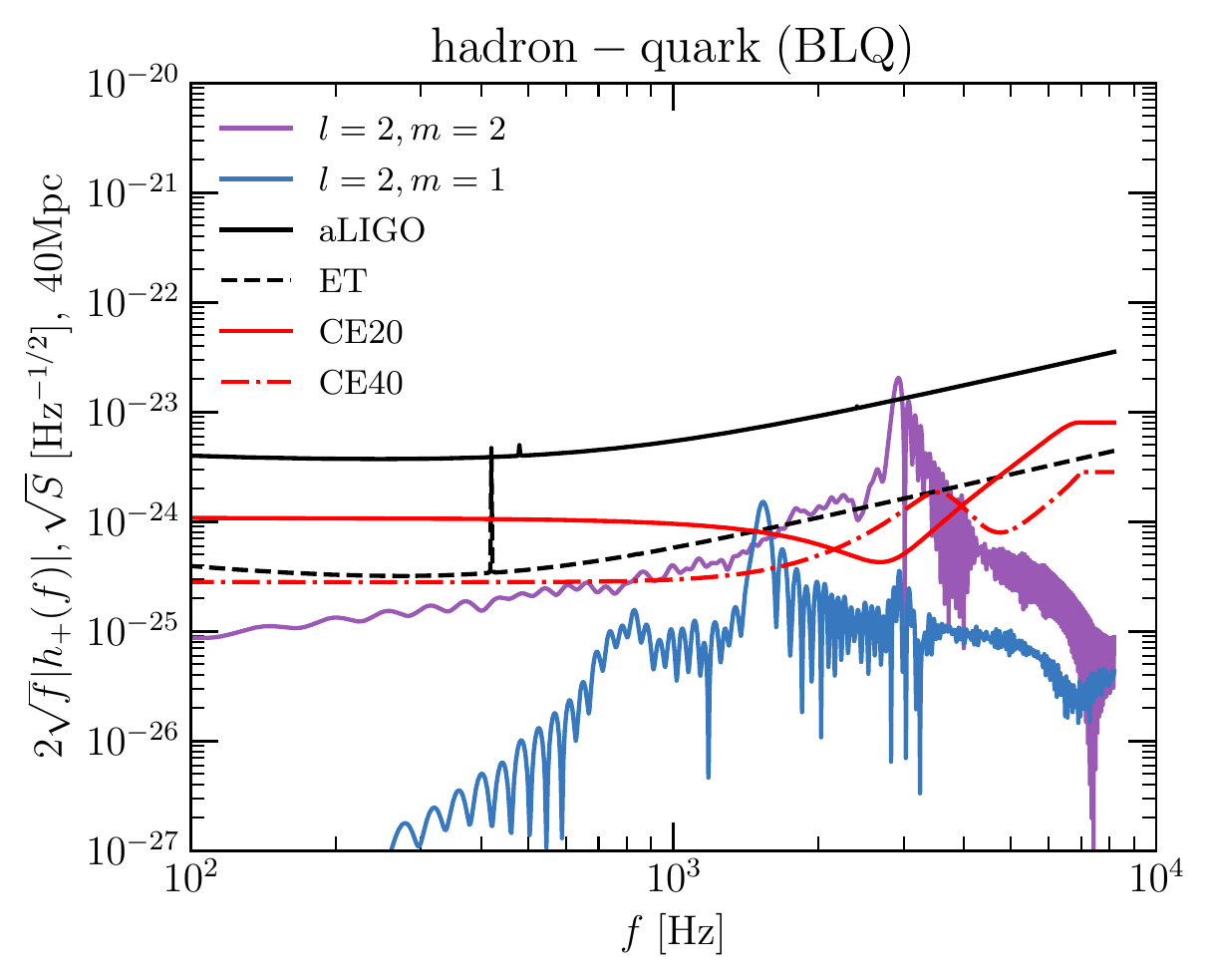}

\caption{ Multimodal GW amplitude spectrum computed for 
symmetric binaries of total mass $\rm{M = 2.6\; M_{\odot}}$ in an edge-on configuration. 
Also shown are the noise sensitivity curves for \rm{advanced LIGO (aLIGO), Einstein 
Telescope (ET)}, the 20 km postmerger-optimized configuration for the Cosmic Explorer 
(CE20) and the 40 km configuration for Cosmic Explorer (CE40). A suppression in the amplitude 
spectral density (ASD)
as a result of the deconfinement phase transition may be detectable with the third 
generation detectors and most cleanly with $\rm{CE40}$.
}
\label{fig:SNR}
\end{figure*}

The one-armed spiral instability 
is a non-axisymmetric mode in a rapidly rotating fluid which, when 
saturated, leads to the dominance of a single high-density mode in the 
fluid density which is displaced from the fluid 
barycenter~\citep{Pickett:1996ApJ, Centrella:2001xp, Saijo:2002nmv, Ou:2006yd}. 
The one-armed spiral instability 
has been observed to develop commonly in BNS merger simulations that 
produce long-lived, massive post-merger remnants 
on timescales of $\mathcal{O}(10 {\rm ms})$~\citep{Paschalidis:2015mla, 
East:2016zvv, PEFS2016, Radice:2016gym, Lehner:2016wjg} and in 
simulations of many other astrophysical systems including 
supernovae~\citep{Ott:2005gj, Kuroda:2013rga}, 
white dwarfs~\citep{Kashyap:2015pog, Kashyap:2017xgn} 
and accretion disks~\citep{Kashyap:2017xgn, Wessel:2020hvu}. 
Each fluid density mode that arises during the evolution of a massive NS 
remnant is associated with GW emission at characteristic frequencies 
stemming from its pattern speed.
As such, the development of the one-armed spiral instability in 
astrophysical systems may be observed by considering multimodal GW 
spectroscopy~\citep{Radice:2016gym}. For the simulations considered in this work we 
extract 
multimodal GW information within the Newman-Penrose formalism. We compute 
the coefficients of $s=-2$ spin-weighted spherical harmonic 
decompositions of the Newman-Penrose scalar 
$\Psi_4$ which we label as $\Psi_4^{l,m}$. The one-armed spiral instability 
can therefore be observed in the GW spectrum extracted from our 
simulations as a growth in the power and amplitude of the $l=2, m=1$ GW 
mode (i.e., $\Psi_4^{2,1}$) and simultaneous decay of the dominant $l=2, m=2$ GW 
mode (i.e., $\Psi_4^{2,2}$). 

Our simulations show that 
\emph{high-density deconfinement phase transitions act to suppress 
the one-armed spiral instability}. Depending on the features of the phase 
transition, the one-armed spiral mode may either arise on a significantly 
longer timescale when compared to simulations which employ purely 
hadronic EOS models, or it may be suppressed altogether on the timescales 
probed by our simulations. 
There are several potential mechanisms via which the instability may be suppressed. 
For example, it has been shown that the physical extent of the remnant plays an 
important role in the development of the instability, with larger remnants being more 
conducive to the development of the instability on shorter 
timescales~\citep{Radice:2016gym, Lehner:2016wjg, Saijo:2016vcd, 
Saijo:2018thy}.
The significant softening at high densities introduced by the phase transition results 
in more compact post-merger remnants (relative to scenarios that consider only 
hadronic degrees of freedom). As such, the more compact hybrid star remnants may see a 
weaker development of the one-armed spiral instability when compared to neutron star 
remnants.

%%Finally, the presence of quarks reduces the moment of inertia of the 
%%remnant by compacting it and therefore, to conserve the angular momentum,
%%the angular velocity increases \dr{But how is this related to the one-armed?}. This effect \dredit{may}{is} also be the mechanism for the increase in 
%%$f_{\rm peak}^{2,2}$ observed in some simulations of BNS mergers with hadron-
%%quark phase transitions.

In the left panel of Fig.~\ref{fig:energy_GW_egap} we show the 
energy carried by the $l=2, m=1$ GW mode (scaled 
by the energy emitted at a time shortly after the merger 
$t_{\rm norm} = t_{\rm mer} + \SI{0.5}{ms}$) as a function of time 
for simulations employing 
the DD2F (hadronic) and DD2F-SF5 (hybrid hadron-quark) EOSs. We find that the energy carried in the $l=2, m=1$ 
mode of the GWs is 
significantly smaller in the simulation employing a hybrid hadron-quark 
EOS, indicating that the one-armed spiral instability is suppressed in 
scenarios with deconfinement phase transitions at densities relevant for 
BNS mergers. We emphasize that in the left panel 
of Fig.~\ref{fig:energy_GW_egap} we showcase results for a set of EOS models which 
are identical below the threshold for a phase transition, and as such the 
simulations have identical initial conditions.
In the right panel of Fig.~\ref{fig:energy_GW_egap}, we show the time-averaged energy 
emitted by the $l=2, m=1$ GW mode $\langle E_{\rm GW}^{2,1} \rangle$ (again scaled 
by the energy emitted at a time shortly after the merger $t_{\rm norm} = t_{\rm mer} + \SI{0.5}{ms}$) as a function of the energy density gap 
$\Delta e$, where we define the energy density gap as the difference 
between the energy density $e$ at the end of the hadronic phase 
and beginning of the quark phase for cold matter 
in $\beta$-equilibrium~\citep{Alford:2015gna},
\begin{equation}\label{eq:energy_gap}
\Delta e \equiv e_{\rm quark, initial} - e_{\rm hadron, final},
\end{equation}
where we assume units where the speed of light $c=1$.
We identify the end and beginning of each phase by considering the change in the 
approximate adiabatic index 
$\Gamma = {\rm d} \log p/{\rm d} \log (\rho)$, where $p$ is the fluid pressure of 
the cold, beta-equilibrium, barotropic EOS for each EOS model considered. The region 
corresponding to the phase transition is always unambiguously marked by 
discontinuities in, or sudden changes in the slope of, the adiabatic index for the EOS models we consider.
For the results depicted in the right panel of Fig.~\ref{fig:energy_GW_egap}, 
we time-average over a window of $\Delta t\approx \SI{40}{ms}$ after the merger 
except for 
cases that lead to a remnant collapse on shorter timescales (in such cases, we 
time-average until the collapse of the NS remnant). 
Additionally, for each simulation, we normalize by a
complementary simulation that uses identical initial data but employs a  
hadronic EOS having the same low-density behavior below the phase transition threshold 
as the hybrid hadron-quark EOS. As such, we depict the point corresponding to all 
hadronic EOS 
simulations with a black square at $\Delta e = 0$. Each simulation is 
time-averaged to the same extent as its complementary hadronic simulation.
\emph{We find an anti-correlation between the energy carried in the 
$l=2,m=1$ GW mode and the size of the 
energy density gap}. In other words, as the size of the energy density gap (and thereby 
the qualitative `strength' of the phase transition) increases, GW 
emission in the $l=2,m=1$ mode decreases, which signifies that the 
one-armed spiral instability is further suppressed for EOS models with 
`stronger' deconfinement phase transitions. We elaborate on
the choice of quantities depicted in Fig.~\ref{fig:energy_GW_egap} in App.~\ref{app:fig1_just}.

\section{Discussion}\label{sec:discussion}
The characteristic frequency associated with peak emission in the 
$l=2, m=1$ GW mode has half the value of that associated with the 
$l=2, m=2$ mode (i.e., $f^{2,1}_{\rm peak} = f^{2,2}_{\rm peak}/2$). 
Observationally, a GW signal would contain information at all contributing 
frequencies. However, the dominant GW emission associated with binary coalescence is 
always expected to be from the $l=2, m=2$ contribution, such that 
$f_{\rm peak} = f^{2,2}_{\rm peak}$. 
Therefore, a potential observational signature of the one-armed spiral 
instability is the growth in power of an incoming GW signal at a 
frequency that is half of the dominant frequency; if it 
develops in the post-merger environment, the one-armed spiral instability 
will continuously power the emission of GWs at $f_{\rm peak}/2$, while 
emission in the dominant $f_{\rm peak}$ decays in time ~\citep{Bernuzzi:2015rla}.

In Fig.~\ref{fig:SNR} we show the post-merger GW amplitude spectrum density (ASD) for a symmetric, 
edge-on binary situated at a distance of $\SI{40}{Mpc}$, which is consistent with the 
luminosity distance observed for 
GW170817 \citep{LIGOScientific:2017vwq}. 
The edge-on configuration is the most 
optimal for the detection of an $m=1$ mode. As expected, we see a relative suppression 
of power in the $m=1$ mode (with respect to the complementing hadronic simulation) with 
the onset of a deconfinement phase transition.
In this realistic configuration, coupled with the $\SI{40}{km}$ Cosmic Explorer 
detector~\citep{Reitze:2019iox}, the appearance of quarks in the post-merger 
remnant results 
in a suppression of the postmerger signal-to-noise ratio (SNR) of 
the $(l=2, m=1)$ mode by a factor of 2, from 2.14 in the hadronic case to 1.08 in 
the hadron-quark case. The GW ASD peak of the $l=2,\;m=1$ mode (between 1-2 
$\rm{kHz}$) and the postmerger ASD peak of the 
$l=2,\;m=2$ mode (between 2-4 $\rm{kHz}$), lie respectively in the most sensitive 
regions of the $\SI{40}{km}$ and the $\SI{20}{km}$ postmerger 
optimized Cosmic Explorer configurations. Our analysis recommends an increase in detector sensitivities in the high-frequency regimes (see also \cite{Zhang:2022yab})
for best possible constraints on deconfinement phase transitions in BNS mergers. 

In this letter we have highlighted, for the first time, that high-density 
deconfinement phase transitions act to suppress the one-armed spiral instability.
We find an anti-correlation between the energy carried in the $l=2,m=1$ GW 
mode and the size of the energy density gap which qualitatively separates the hadronic 
and quark phases. Our findings reveal a deep connection between 
observable multimodal GW emission and the microphysical description of 
matter in the post-merger environment. We expect the one-armed spiral 
instability to be detectable at distances of $\SI{40}{Mpc}$ using future generation 
detectors~\citep{Radice:2016gym}.
If evidence of a strong one-armed 
spiral mode can be inferred from GW observations of the post-BNS merger 
environment, our findings suggest that a strong high-density 
deconfinement phase transition at the densities relevant to BNS mergers would be
disfavored. On the other 
hand, if evidence for the one-armed spiral instability 
is \emph{not} found for close-by BNS mergers, this could also point to the 
possibility of a deconfinement phase transition taking place at densities relevant 
to BNS mergers.

We point out that other effects relevant in the 
post-merger environment - such as the presence of strong magnetic 
fields~\citep{Franci:2013mma}
and additional degrees of freedom that can cause a sudden softening of the EOS
- may 
affect the development of the one-armed spiral 
instability. However, the relevant timescales 
and extent to which the aforementioned phenomena can affect the development of 
non-axisymmetric instabilities or the GW spectrum 
remains uncertain~\citep{Radice:2016gym, Muhlberger:2014pja}, and may not impact 
our conclusions~\citep{Palenzuela:2021gdo, Zappa:2022rpd}.
%%\dr{I think this 
%%sentence has a logic problem: we need to consider these effects if only to make 
%%sure they are not important. However, I would turn this around and argue on the 
%%basis of Palenzuela and Zappa that, as far as we know now, these effects will 
%%not impact our conclusions.}. 
The 
effects discussed in the 
present work arise on dynamical timescales $\sim \mathcal{O}(\SI{10}{ms})$, and may 
be the dominant mechanism for suppression of the one-armed spiral instability. 
Additionally, although we find a trend in the decrease of energy carried by the 
$l=2,m=1$ GW mode for larger values of $\Delta e$, additional studies will help 
establish a more robust trend and provide an understanding of the potential spread in 
the trend. In particular, future lines of investigation will include: (1) considering 
the combined effects of the mass ratio and high-density phase transitions on the 
development of the one-armed spiral instability; (2) considering the effects of 
accurate neutrino transport on high-density deconfinement phase transitions, as 
neutrinos may modify the composition of matter and thereby potentially 
affect the onset of the phase transition; (3) employing EOS models at 
systematically increasing values of $\Delta e$ while holding the hadronic region of 
the EOS fixed, as a limitation of the present work is the assumption that the 
$l=2, m=1$ GW mode is perfectly known in the case of hadronic EOSs; and (4) 
investigating the effects discussed in this work in scenarios with 
a \emph{crossover} to 
quark matter, as our present work only considers EOS models with phase 
transitions.
We leave such studies to future work.\\

%%\begin{acknowledgements}
%%\paragraph{Acknowledgements.}
PE acknowledges funding from the National Science Foundation under Grant
No. PHY-2020275.
DR acknowledges funding from the U.S. Department of Energy, Office of
Science, Division of Nuclear Physics under Award Number(s) DE-SC0021177
and from the National Science Foundation under Grants No. PHY-2011725,
PHY-2116686, and AST-2108467.
Simulations were performed on Bridges2 and Expanse (NSF XSEDE allocation
TG-PHY160025). 
This research used resources of the National Energy Research Scientific
Computing Center, a DOE Office of Science User Facility supported by the
Office of Science of the U.S.~Department of Energy under Contract
No.~DE-AC02-05CH11231.
%\end{acknowledgements}

\bibliography{references.bib}

\begin{thebibliography}{}
\expandafter\ifx\csname natexlab\endcsname\relax\def\natexlab#1{#1}\fi
\providecommand{\url}[1]{\href{#1}{#1}}
\providecommand{\dodoi}[1]{doi:~\href{http://doi.org/#1}{\nolinkurl{#1}}}
\providecommand{\doeprint}[1]{\href{http://ascl.net/#1}{\nolinkurl{http://ascl.net/#1}}}
\providecommand{\doarXiv}[1]{\href{https://arxiv.org/abs/#1}{\nolinkurl{https://arxiv.org/abs/#1}}}

\bibitem[{Abbott {et~al.}(2017{\natexlab{a}})}]{LIGOScientific:2017fdd}
Abbott, B.~P., {et~al.} 2017{\natexlab{a}}, Astrophys. J. Lett., 851, L16,
  \dodoi{10.3847/2041-8213/aa9a35}

\bibitem[{Abbott {et~al.}(2017{\natexlab{b}})}]{LIGOScientific:2017vwq}
---. 2017{\natexlab{b}}, Phys. Rev. Lett., 119, 161101

\bibitem[{Abbott {et~al.}(2018)}]{LIGOScientific:2018cki}
---. 2018, Phys. Rev. Lett., 121, 161101,
  \dodoi{10.1103/PhysRevLett.121.161101}

\bibitem[{Abbott {et~al.}(2019)}]{LIGOScientific:2018hze}
---. 2019, Phys. Rev. X, 9, 011001, \dodoi{10.1103/PhysRevX.9.011001}

\bibitem[{Ackley {et~al.}(2020)}]{Ackley:2020atn}
Ackley, K., {et~al.} 2020, Publ. Astron. Soc. Austral., 37, e047,
  \dodoi{10.1017/pasa.2020.39}

\bibitem[{Alford \& Sedrakian(2017)}]{T92}
Alford, M., \& Sedrakian, A. 2017, Phys. Rev. Lett., 119, 161104,
  \dodoi{10.1103/PhysRevLett.119.161104}

\bibitem[{Alford \& Han(2016)}]{Alford:2015gna}
Alford, M.~G., \& Han, S. 2016, Eur. Phys. J. A, 52, 62,
  \dodoi{10.1140/epja/i2016-16062-9}

\bibitem[{Alvarez-Castillo \& Blaschke(2017)}]{T91}
Alvarez-Castillo, D.~E., \& Blaschke, D.~B. 2017, Phys. Rev. C, 96, 045809,
  \dodoi{10.1103/PhysRevC.96.045809}

\bibitem[{Annala {et~al.}(2022)Annala, Gorda, Katerini, Kurkela, N\"attil\"a,
  Paschalidis, \& Vuorinen}]{Annala:2021gom}
Annala, E., Gorda, T., Katerini, E., {et~al.} 2022, Phys. Rev. X, 12, 011058,
  \dodoi{10.1103/PhysRevX.12.011058}

\bibitem[{Bastian(2021)}]{PhysRevD.103.023001}
Bastian, N.-U.~F. 2021, Phys. Rev. D, 103, 023001,
  \dodoi{10.1103/PhysRevD.103.023001}

\bibitem[{Bauswein {et~al.}(2019)Bauswein, Bastian, Blaschke, Chatziioannou,
  Clark, Fischer, \& Oertel}]{Bauswein:2018bma}
Bauswein, A., Bastian, N.-U.~F., Blaschke, D.~B., {et~al.} 2019, Phys. Rev.
  Lett., 122, 061102, \dodoi{10.1103/PhysRevLett.122.061102}

\bibitem[{Bauswein {et~al.}(2013)Bauswein, Baumgarte, \&
  Janka}]{Bauswein:2013jpa}
Bauswein, A., Baumgarte, T.~W., \& Janka, H.~T. 2013, Phys. Rev. Lett., 111,
  131101, \dodoi{10.1103/PhysRevLett.111.131101}

\bibitem[{Bauswein {et~al.}(2021)Bauswein, Blacker, Lioutas, Soultanis,
  Vijayan, \& Stergioulas}]{Bauswein:2020xlt}
Bauswein, A., Blacker, S., Lioutas, G., {et~al.} 2021, Phys. Rev. D, 103,
  123004, \dodoi{10.1103/PhysRevD.103.123004}

\bibitem[{Bauswein {et~al.}(2017)Bauswein, Just, Janka, \&
  Stergioulas}]{Bauswein:2017vtn}
Bauswein, A., Just, O., Janka, H.-T., \& Stergioulas, N. 2017, Astrophys. J.
  Lett., 850, L34, \dodoi{10.3847/2041-8213/aa9994}

\bibitem[{Bauswein {et~al.}(2020)Bauswein, Blacker, Vijayan, Stergioulas,
  Chatziioannou, Clark, Bastian, Blaschke, Cierniak, \&
  Fischer}]{Bauswein:2020aag}
Bauswein, A., Blacker, S., Vijayan, V., {et~al.} 2020, Phys. Rev. Lett., 125,
  141103, \dodoi{10.1103/PhysRevLett.125.141103}

\bibitem[{Bernuzzi {et~al.}(2015)Bernuzzi, Dietrich, \&
  Nagar}]{Bernuzzi:2015rla}
Bernuzzi, S., Dietrich, T., \& Nagar, A. 2015, Phys. Rev. Lett., 115, 091101,
  \dodoi{10.1103/PhysRevLett.115.091101}

\bibitem[{Blacker {et~al.}(2020)Blacker, Bastian, Bauswein, Blaschke, Fischer,
  Oertel, Soultanis, \& Typel}]{Blacker:2020nlq}
Blacker, S., Bastian, N.-U.~F., Bauswein, A., {et~al.} 2020, Phys. Rev. D, 102,
  123023, \dodoi{10.1103/PhysRevD.102.123023}

\bibitem[{Bombaci \& Logoteta(2018)}]{Bombaci:2018ksa}
Bombaci, I., \& Logoteta, D. 2018, Astron. Astrophys., 609, A128,
  \dodoi{10.1051/0004-6361/201731604}

\bibitem[{Bozzola {et~al.}(2019)Bozzola, Espino, Lewin, \&
  Paschalidis}]{Bozzola:2019tit}
Bozzola, G., Espino, P.~L., Lewin, C.~D., \& Paschalidis, V. 2019.
\newblock \doarXiv{1905.00028}

\bibitem[{Breschi {et~al.}(2022)Breschi, Bernuzzi, Godzieba, Perego, \&
  Radice}]{Breschi:2021xrx}
Breschi, M., Bernuzzi, S., Godzieba, D., Perego, A., \& Radice, D. 2022, Phys.
  Rev. Lett., 128, 161102, \dodoi{10.1103/PhysRevLett.128.161102}

\bibitem[{Breschi {et~al.}(2021)Breschi, Perego, Bernuzzi, Del~Pozzo, Nedora,
  Radice, \& Vescovi}]{Breschi:2021tbm}
Breschi, M., Perego, A., Bernuzzi, S., {et~al.} 2021, Mon. Not. Roy. Astron.
  Soc., 505, 1661, \dodoi{10.1093/mnras/stab1287}

\bibitem[{Capano {et~al.}(2020)Capano, Tews, Brown, Margalit, De, Kumar, Brown,
  Krishnan, \& Reddy}]{Capano:2019eae}
Capano, C.~D., Tews, I., Brown, S.~M., {et~al.} 2020, Nature Astron., 4, 625,
  \dodoi{10.1038/s41550-020-1014-6}

\bibitem[{Centrella {et~al.}(2001)Centrella, New, Lowe, \&
  Brown}]{Centrella:2001xp}
Centrella, J.~M., New, K. C.~B., Lowe, L.~L., \& Brown, J.~D. 2001, Astrophys.
  J. Lett., 550, L193, \dodoi{10.1086/319634}

\bibitem[{Coughlin {et~al.}(2019)Coughlin, Dietrich, Margalit, \&
  Metzger}]{Coughlin:2018fis}
Coughlin, M.~W., Dietrich, T., Margalit, B., \& Metzger, B.~D. 2019, Mon. Not.
  Roy. Astron. Soc., 489, L91, \dodoi{10.1093/mnrasl/slz133}

\bibitem[{Damour {et~al.}(2012)Damour, Nagar, \& Villain}]{Damour:2012yf}
Damour, T., Nagar, A., \& Villain, L. 2012, Phys. Rev. D, 85, 123007,
  \dodoi{10.1103/PhysRevD.85.123007}

\bibitem[{De {et~al.}(2018)De, Finstad, Lattimer, Brown, Berger, \&
  Biwer}]{De:2018uhw}
De, S., Finstad, D., Lattimer, J.~M., {et~al.} 2018, Phys. Rev. Lett., 121,
  091102, \dodoi{10.1103/PhysRevLett.121.091102}

\bibitem[{Dietrich {et~al.}(2020)Dietrich, Coughlin, Pang, Bulla, Heinzel,
  Issa, Tews, \& Antier}]{Dietrich:2020efo}
Dietrich, T., Coughlin, M.~W., Pang, P. T.~H., {et~al.} 2020, Science, 370,
  1450, \dodoi{10.1126/science.abb4317}

\bibitem[{East {et~al.}(2016)East, Paschalidis, \& Pretorius}]{East:2016zvv}
East, W.~E., Paschalidis, V., \& Pretorius, F. 2016, Class. Quant. Grav., 33,
  244004, \dodoi{10.1088/0264-9381/33/24/244004}

\bibitem[{{East} {et~al.}(2016){East}, {Paschalidis}, {Pretorius}, \&
  {Shapiro}}]{PEFS2016}
{East}, W.~E., {Paschalidis}, V., {Pretorius}, F., \& {Shapiro}, S.~L. 2016,
  prd, 93, 024011, \dodoi{10.1103/PhysRevD.93.024011}

\bibitem[{Espino \& Paschalidis(2022)}]{Espino:2021adh}
Espino, P.~L., \& Paschalidis, V. 2022, Phys. Rev. D, 105, 043014,
  \dodoi{10.1103/PhysRevD.105.043014}

\bibitem[{Espino {et~al.}(2019)Espino, Paschalidis, Baumgarte, \&
  Shapiro}]{Espino:2019xcl}
Espino, P.~L., Paschalidis, V., Baumgarte, T.~W., \& Shapiro, S.~L. 2019, Phys.
  Rev. D, 100, 043014, \dodoi{10.1103/PhysRevD.100.043014}

\bibitem[{Evans {et~al.}(2021)}]{Evans:2021gyd}
Evans, M., {et~al.} 2021, arXiv.
\newblock \doarXiv{2109.09882}

\bibitem[{Franci {et~al.}(2013)Franci, De~Pietri, Dionysopoulou, \&
  Rezzolla}]{Franci:2013mma}
Franci, L., De~Pietri, R., Dionysopoulou, K., \& Rezzolla, L. 2013, Phys. Rev.
  D, 88, 104028, \dodoi{10.1103/PhysRevD.88.104028}

\bibitem[{Fujimoto {et~al.}(2022)Fujimoto, Fukushima, Hotokezaka, \&
  Kyutoku}]{Fujimoto:2022xhv}
Fujimoto, Y., Fukushima, K., Hotokezaka, K., \& Kyutoku, K. 2022, arXiv.
\newblock \doarXiv{2205.03882}

\bibitem[{Gourgoulhon {et~al.}(2001)Gourgoulhon, Grandclement, Taniguchi,
  Marck, \& Bonazzola}]{Gourgoulhon:2000nn}
Gourgoulhon, E., Grandclement, P., Taniguchi, K., Marck, J.-A., \& Bonazzola,
  S. 2001, Phys. Rev. D, 63, 064029, \dodoi{10.1103/PhysRevD.63.064029}

\bibitem[{Hinderer {et~al.}(2010)Hinderer, Lackey, Lang, \&
  Read}]{Hinderer:2009ca}
Hinderer, T., Lackey, B.~D., Lang, R.~N., \& Read, J.~S. 2010, Phys. Rev. D,
  81, 123016, \dodoi{10.1103/PhysRevD.81.123016}

\bibitem[{Hotokezaka {et~al.}(2011)Hotokezaka, Kyutoku, Okawa, Shibata, \&
  Kiuchi}]{Hotokezaka:2011dh}
Hotokezaka, K., Kyutoku, K., Okawa, H., Shibata, M., \& Kiuchi, K. 2011, Phys.
  Rev. D, 83, 124008, \dodoi{10.1103/PhysRevD.83.124008}

\bibitem[{Kashyap {et~al.}(2015)Kashyap, Fisher, Garc\'\i{}a-Berro,
  Aznar-Sigu\'an, Ji, \& Lor\'en-Aguilar}]{Kashyap:2015pog}
Kashyap, R., Fisher, R., Garc\'\i{}a-Berro, E., {et~al.} 2015, Astrophys. J.,
  800, L7, \dodoi{10.1088/2041-8205/800/1/l7}

\bibitem[{Kashyap {et~al.}(2017)Kashyap, Fisher, Garc\'\i{}a-Berro,
  Aznar-Sigu\'an, Ji, \& Lor\'en-Aguilar}]{Kashyap:2017xgn}
---. 2017, Astrophys. J., 840, 16, \dodoi{10.3847/1538-4357/aa6afb}

\bibitem[{Kashyap {et~al.}(2022)}]{Kashyap:2021wzs}
Kashyap, R., {et~al.} 2022, Phys. Rev. D, 105, 103022,
  \dodoi{10.1103/PhysRevD.105.103022}

\bibitem[{Kedia {et~al.}(2022)Kedia, Kim, Suh, \& Mathews}]{Kedia:2022nns}
Kedia, A., Kim, H.~I., Suh, I.-S., \& Mathews, G.~J. 2022, arXiv.
\newblock \doarXiv{2203.05461}

\bibitem[{Kuroda {et~al.}(2014)Kuroda, Takiwaki, \& Kotake}]{Kuroda:2013rga}
Kuroda, T., Takiwaki, T., \& Kotake, K. 2014, Phys. Rev. D, 89, 044011,
  \dodoi{10.1103/PhysRevD.89.044011}

\bibitem[{Lee {et~al.}(2010)Lee, Ramirez-Ruiz, \& van~de Ven}]{Lee:2009ca}
Lee, W.~H., Ramirez-Ruiz, E., \& van~de Ven, G. 2010, Astrophys. J., 720, 953,
  \dodoi{10.1088/0004-637X/720/1/953}

\bibitem[{Lehner {et~al.}(2016)Lehner, Liebling, Palenzuela, \&
  Motl}]{Lehner:2016wjg}
Lehner, L., Liebling, S.~L., Palenzuela, C., \& Motl, P.~M. 2016, Phys. Rev.,
  D94, 043003, \dodoi{10.1103/PhysRevD.94.043003}

\bibitem[{Liebling {et~al.}(2021)Liebling, Palenzuela, \&
  Lehner}]{Liebling:2020dhf}
Liebling, S.~L., Palenzuela, C., \& Lehner, L. 2021, Class. Quant. Grav., 38,
  115007, \dodoi{10.1088/1361-6382/abf898}

\bibitem[{Loffredo {et~al.}(2022)Loffredo, Perego, Logoteta, \&
  Branchesi}]{Loffredo:2022prq}
Loffredo, E., Perego, A., Logoteta, D., \& Branchesi, M. 2022.
\newblock \doarXiv{2209.04458}

\bibitem[{Logoteta(2021)}]{Logoteta:2021iuy}
Logoteta, D. 2021, Universe, 7, 408, \dodoi{10.3390/universe7110408}

\bibitem[{Logoteta {et~al.}(2021)Logoteta, Perego, \&
  Bombaci}]{Logoteta:2020yxf}
Logoteta, D., Perego, A., \& Bombaci, I. 2021, Astron. Astrophys., 646, A55,
  \dodoi{10.1051/0004-6361/202039457}

\bibitem[{Lovato {et~al.}(2022)}]{Lovato:2022vgq}
Lovato, A., {et~al.} 2022.
\newblock \doarXiv{2211.02224}

\bibitem[{Margalit \& Metzger(2017)}]{Margalit:2017dij}
Margalit, B., \& Metzger, B.~D. 2017, Astrophys. J. Lett., 850, L19,
  \dodoi{10.3847/2041-8213/aa991c}

\bibitem[{Most {et~al.}(2020)Most, Jens~Papenfort, Dexheimer, Hanauske,
  Stoecker, \& Rezzolla}]{Most:2019onn}
Most, E.~R., Jens~Papenfort, L., Dexheimer, V., {et~al.} 2020, Eur. Phys. J. A,
  56, 59

\bibitem[{Most {et~al.}(2019)Most, Papenfort, Dexheimer, Hanauske, Schramm,
  St\"ocker, \& Rezzolla}]{Most:2018eaw}
Most, E.~R., Papenfort, L.~J., Dexheimer, V., {et~al.} 2019, Phys. Rev. Lett.,
  122, 061101, \dodoi{10.1103/PhysRevLett.122.061101}

\bibitem[{Muhlberger {et~al.}(2014)Muhlberger, Nouri, Duez, Foucart, Kidder,
  Ott, Scheel, Szil\'agyi, \& Teukolsky}]{Muhlberger:2014pja}
Muhlberger, C.~D., Nouri, F.~H., Duez, M.~D., {et~al.} 2014, Phys. Rev. D, 90,
  104014, \dodoi{10.1103/PhysRevD.90.104014}

\bibitem[{Ott {et~al.}(2005)Ott, Ou, Tohline, \& Burrows}]{Ott:2005gj}
Ott, C.~D., Ou, S., Tohline, J.~E., \& Burrows, A. 2005, Astrophys. J. Lett.,
  625, L119, \dodoi{10.1086/431305}

\bibitem[{Ou \& Tohline(2006)}]{Ou:2006yd}
Ou, S., \& Tohline, J. 2006, Astrophys. J., 651, 1068, \dodoi{10.1086/507597}

\bibitem[{Palenzuela {et~al.}(2022)Palenzuela, Aguilera-Miret, Carrasco,
  Ciolfi, Kalinani, Kastaun, Mi\~nano, \& Vigan\`o}]{Palenzuela:2021gdo}
Palenzuela, C., Aguilera-Miret, R., Carrasco, F., {et~al.} 2022, Phys. Rev. D,
  106, 023013, \dodoi{10.1103/PhysRevD.106.023013}

\bibitem[{Pang {et~al.}(2021)Pang, Tews, Coughlin, Bulla, Van Den~Broeck, \&
  Dietrich}]{Pang:2021jta}
Pang, P. T.~H., Tews, I., Coughlin, M.~W., {et~al.} 2021, Astrophys. J., 922,
  14, \dodoi{10.3847/1538-4357/ac19ab}

\bibitem[{Paschalidis {et~al.}(2015)Paschalidis, East, Pretorius, \&
  Shapiro}]{Paschalidis:2015mla}
Paschalidis, V., East, W.~E., Pretorius, F., \& Shapiro, S.~L. 2015, Phys.
  Rev., D92, 121502, \dodoi{10.1103/PhysRevD.92.121502}

\bibitem[{Paschalidis {et~al.}(2018)Paschalidis, Yagi, Alvarez-Castillo,
  Blaschke, \& Sedrakian}]{Paschalidis:2017qmb}
Paschalidis, V., Yagi, K., Alvarez-Castillo, D., Blaschke, D.~B., \& Sedrakian,
  A. 2018, Phys. Rev., D97, 084038, \dodoi{10.1103/PhysRevD.97.084038}

\bibitem[{Perego {et~al.}(2019)Perego, Bernuzzi, \& Radice}]{Perego:2019adq}
Perego, A., Bernuzzi, S., \& Radice, D. 2019, Eur. Phys. J. A, 55, 124,
  \dodoi{10.1140/epja/i2019-12810-7}

\bibitem[{Perego {et~al.}(2022)Perego, Logoteta, Radice, Bernuzzi, Kashyap,
  Das, Padamata, \& Prakash}]{Perego:2021mkd}
Perego, A., Logoteta, D., Radice, D., {et~al.} 2022, Phys. Rev. Lett., 129,
  032701, \dodoi{10.1103/PhysRevLett.129.032701}

\bibitem[{{Pickett} {et~al.}(1996){Pickett}, {Durisen}, \&
  {Davis}}]{Pickett:1996ApJ}
{Pickett}, B.~K., {Durisen}, R.~H., \& {Davis}, G.~A. 1996, \apj, 458, 714,
  \dodoi{10.1086/176852}

\bibitem[{Pollney {et~al.}(2011)Pollney, Reisswig, Schnetter, Dorband, \&
  Diener}]{Pollney:2009yz}
Pollney, D., Reisswig, C., Schnetter, E., Dorband, N., \& Diener, P. 2011,
  Phys. Rev. D, 83, 044045, \dodoi{10.1103/PhysRevD.83.044045}

\bibitem[{Prakash {et~al.}(2021)Prakash, Radice, Logoteta, Perego, Nedora,
  Bombaci, Kashyap, Bernuzzi, \& Endrizzi}]{Prakash:2021wpz}
Prakash, A., Radice, D., Logoteta, D., {et~al.} 2021, Phys. Rev. D, 104,
  083029, \dodoi{10.1103/PhysRevD.104.083029}

\bibitem[{Punturo {et~al.}(2010)}]{Punturo:2010zz}
Punturo, M., {et~al.} 2010, Class. Quant. Grav., 27, 194002,
  \dodoi{10.1088/0264-9381/27/19/194002}

\bibitem[{Radice {et~al.}(2017)Radice, Bernuzzi, Del~Pozzo, Roberts, \&
  Ott}]{Radice:2016rys}
Radice, D., Bernuzzi, S., Del~Pozzo, W., Roberts, L.~F., \& Ott, C.~D. 2017,
  Astrophys. J. Lett., 842, L10, \dodoi{10.3847/2041-8213/aa775f}

\bibitem[{Radice {et~al.}(2016{\natexlab{a}})Radice, Bernuzzi, \&
  Ott}]{Radice:2016gym}
Radice, D., Bernuzzi, S., \& Ott, C.~D. 2016{\natexlab{a}}, Phys. Rev., D94,
  064011, \dodoi{10.1103/PhysRevD.94.064011}

\bibitem[{Radice {et~al.}(2020)Radice, Bernuzzi, \& Perego}]{Radice:2020ddv}
Radice, D., Bernuzzi, S., \& Perego, A. 2020, Ann. Rev. Nucl. Part. Sci., 70,
  95, \dodoi{10.1146/annurev-nucl-013120-114541}

\bibitem[{Radice {et~al.}(2022)Radice, Bernuzzi, Perego, \&
  Haas}]{Radice:2021jtw}
Radice, D., Bernuzzi, S., Perego, A., \& Haas, R. 2022, Mon. Not. Roy. Astron.
  Soc., 512, 1499, \dodoi{10.1093/mnras/stac589}

\bibitem[{Radice \& Dai(2019)}]{Radice:2018ozg}
Radice, D., \& Dai, L. 2019, Eur. Phys. J. A, 55, 50,
  \dodoi{10.1140/epja/i2019-12716-4}

\bibitem[{Radice {et~al.}(2016{\natexlab{b}})Radice, Galeazzi, Lippuner,
  Roberts, Ott, \& Rezzolla}]{Radice:2016dwd}
Radice, D., Galeazzi, F., Lippuner, J., {et~al.} 2016{\natexlab{b}}, Mon. Not.
  Roy. Astron. Soc., 460, 3255, \dodoi{10.1093/mnras/stw1227}

\bibitem[{Radice {et~al.}(2018{\natexlab{a}})Radice, Perego, Hotokezaka, Fromm,
  Bernuzzi, \& Roberts}]{Radice:2018pdn}
Radice, D., Perego, A., Hotokezaka, K., {et~al.} 2018{\natexlab{a}}, Astrophys.
  J., 869, 130, \dodoi{10.3847/1538-4357/aaf054}

\bibitem[{Radice {et~al.}(2018{\natexlab{b}})Radice, Perego, Zappa, \&
  Bernuzzi}]{Radice:2017lry}
Radice, D., Perego, A., Zappa, F., \& Bernuzzi, S. 2018{\natexlab{b}},
  Astrophys. J. Lett., 852, L29, \dodoi{10.3847/2041-8213/aaa402}

\bibitem[{Radice \& Rezzolla(2012)}]{Radice:2012cu}
Radice, D., \& Rezzolla, L. 2012, Astron. Astrophys., 547, A26,
  \dodoi{10.1051/0004-6361/201219735}

\bibitem[{Radice {et~al.}(2014{\natexlab{a}})Radice, Rezzolla, \&
  Galeazzi}]{Radice:2013hxh}
Radice, D., Rezzolla, L., \& Galeazzi, F. 2014{\natexlab{a}}, Mon. Not. Roy.
  Astron. Soc., 437, L46, \dodoi{10.1093/mnrasl/slt137}

\bibitem[{Radice {et~al.}(2014{\natexlab{b}})Radice, Rezzolla, \&
  Galeazzi}]{Radice:2013xpa}
---. 2014{\natexlab{b}}, Class. Quant. Grav., 31, 075012,
  \dodoi{10.1088/0264-9381/31/7/075012}

\bibitem[{Read {et~al.}(2009)Read, Lackey, Owen, \& Friedman}]{Read2009}
Read, J.~S., Lackey, B.~D., Owen, B.~J., \& Friedman, J.~L. 2009, Phys. Rev. D,
  79, 124032, \dodoi{10.1103/PhysRevD.79.124032}

\bibitem[{Reitze {et~al.}(2019)}]{Reitze:2019iox}
Reitze, D., {et~al.} 2019, Bull. Am. Astron. Soc., 51, 035.
\newblock \doarXiv{1907.04833}

\bibitem[{Saijo(2018)}]{Saijo:2018thy}
Saijo, M. 2018, Phys. Rev. D, 98, 024003

\bibitem[{Saijo {et~al.}(2002)Saijo, Baumgarte, \& Shapiro}]{Saijo:2002nmv}
Saijo, M., Baumgarte, T.~W., \& Shapiro, S.~L. 2002, Astrophys. J., 595, 352,
  \dodoi{10.1086/377334}

\bibitem[{Saijo \& Yoshida(2016)}]{Saijo:2016vcd}
Saijo, M., \& Yoshida, S. 2016, Phys. Rev. D, 94, 084032

\bibitem[{Schnetter {et~al.}(2004)Schnetter, Hawley, \&
  Hawke}]{Schnetter:2003rb}
Schnetter, E., Hawley, S.~H., \& Hawke, I. 2004, Class. Quant. Grav., 21, 1465,
  \dodoi{10.1088/0264-9381/21/6/014}

\bibitem[{Sekiguchi {et~al.}(2011)Sekiguchi, Kiuchi, Kyutoku, \&
  Shibata}]{Sekiguchi:2011mc}
Sekiguchi, Y., Kiuchi, K., Kyutoku, K., \& Shibata, M. 2011, Phys. Rev. Lett.,
  107, 211101, \dodoi{10.1103/PhysRevLett.107.211101}

\bibitem[{Shibata(2005)}]{Shibata:2005xz}
Shibata, M. 2005, Phys. Rev. Lett., 94, 201101,
  \dodoi{10.1103/PhysRevLett.94.201101}

\bibitem[{Taniguchi \& Gourgoulhon(2002)}]{Taniguchi:2001ww}
Taniguchi, K., \& Gourgoulhon, E. 2002, Phys. Rev. D, 65, 044027,
  \dodoi{10.1103/PhysRevD.65.044027}

\bibitem[{Taniguchi {et~al.}(2001)Taniguchi, Gourgoulhon, \&
  Bonazzola}]{Taniguchi:2001qv}
Taniguchi, K., Gourgoulhon, E., \& Bonazzola, S. 2001, Phys. Rev. D, 64,
  064012, \dodoi{10.1103/PhysRevD.64.064012}

\bibitem[{Tootle {et~al.}(2022)Tootle, Ecker, Topolski, Demircik, Jarvinen, \&
  Rezzolla}]{Tootle:2022pvd}
Tootle, S., Ecker, C., Topolski, K., {et~al.} 2022, arXiv.
\newblock \doarXiv{2205.05691}

\bibitem[{Weih {et~al.}(2020)Weih, Hanauske, \& Rezzolla}]{Weih:2019xvw}
Weih, L.~R., Hanauske, M., \& Rezzolla, L. 2020, Phys. Rev. Lett., 124, 171103,
  \dodoi{10.1103/PhysRevLett.124.171103}

\bibitem[{Wessel {et~al.}(2021)Wessel, Paschalidis, Tsokaros, Ruiz, \&
  Shapiro}]{Wessel:2020hvu}
Wessel, E., Paschalidis, V., Tsokaros, A., Ruiz, M., \& Shapiro, S.~L. 2021,
  Phys. Rev. D, 103, 043013, \dodoi{10.1103/PhysRevD.103.043013}

\bibitem[{York(1999)}]{York:1998hy}
York, Jr., J.~W. 1999, Phys. Rev. Lett., 82, 1350,
  \dodoi{10.1103/PhysRevLett.82.1350}

\bibitem[{Zappa {et~al.}(2022)Zappa, Bernuzzi, Radice, \&
  Perego}]{Zappa:2022rpd}
Zappa, F., Bernuzzi, S., Radice, D., \& Perego, A. 2022.
\newblock \doarXiv{2210.11491}

\bibitem[{Zhang {et~al.}(2022)Zhang, Yang, Martynov, Schmidt, \&
  Miao}]{Zhang:2022yab}
Zhang, T., Yang, H., Martynov, D., Schmidt, P., \& Miao, H. 2022.
\newblock \doarXiv{2212.12144}

\bibitem[{Zlochower {et~al.}(2022)Zlochower, Brandt, Diener, Gabella,
  Gracia-Linares, Haas, Kedia, Alcubierre, Alic, Allen, Ansorg,
  Babiuc-Hamilton, Baiotti, Benger, Bentivegna, Bernuzzi, Bode, Bozzola,
  Brendal, Bruegmann, Campanelli, Cipolletta, Corvino, Cupp, Pietri,
  Dimmelmeier, Dooley, Dorband, Elley, Khamra, Etienne, Faber, Font, Frieben,
  Giacomazzo, Goodale, Gundlach, Hawke, Hawley, Hinder, Huerta, Husa, Iyer,
  Johnson, Joshi, Kastaun, Kellermann, Knapp, Koppitz, Laguna, Lanferman,
  L{\"o}ffler, Masso, Menger, Merzky, Miller, Miller, Moesta, Montero, Mundim,
  Nelson, Nerozzi, Noble, Ott, Paruchuri, Pollney, Radice, Radke, Reisswig,
  Rezzolla, Rideout, Ripeanu, Sala, Schewtschenko, Schnetter, Schutz, Seidel,
  Seidel, Shalf, Sible, Sperhake, Stergioulas, Suen, Szilagyi, Takahashi,
  Thomas, Thornburg, Tobias, Tonita, Walker, Wan, Wardell, Werneck, Witek,
  Zilh{\~a}o, \& Zink}]{EinsteinToolkit:2022_05}
Zlochower, Y., Brandt, S.~R., Diener, P., {et~al.} 2022, The Einstein Toolkit,
  The "Berhard Riemann" release, ET\_2022\_05,  Zenodo,
  \dodoi{10.5281/zenodo.6588641}

\end{thebibliography}
\acrodef{ADM}{Arnowitt-Deser-Misner}
\acrodef{AMR}{adaptive mesh-refinement}
\acrodef{BH}{black hole}
\acrodef{BBH}{binary black-hole}
\acrodef{BHNS}{black-hole neutron-star}
\acrodef{BNS}{binary neutron star}
\acrodef{CCSN}{core-collapse supernova}
\acrodefplural{CCSN}[CCSNe]{core-collapse supernovae}
\acrodef{CMA}{consistent multi-fluid advection}
\acrodef{CFL}{Courant-Friedrichs-Lewy}
\acrodef{DG}{discontinuous Galerkin}
\acrodef{HMNS}{hypermassive neutron star}
\acrodef{EM}{electromagnetic}
\acrodef{ET}{Einstein Telescope}
\acrodef{EOB}{effective-one-body}
\acrodef{EOS}{equation of state}
\acrodef{FF}{fitting factor}
\acrodef{GR}{general-relativistic}
\acrodef{GRLES}{general-relativistic large-eddy simulation}
\acrodef{GRHD}{general-relativistic hydrodynamics}
\acrodef{GRMHD}{general-relativistic magnetohydrodynamics}
\acrodef{GW}{gravitational wave}
\acrodef{ILES}{implicit large-eddy simulations}
\acrodef{LIA}{linear interaction analysis}
\acrodef{LES}{large-eddy simulation}
\acrodefplural{LES}[LES]{large-eddy simulations}
\acrodef{MRI}{magnetorotational instability}
\acrodef{NR}{numerical relativity}
\acrodef{NS}{neutron star}
\acrodef{PNS}{protoneutron star}
\acrodef{RMNS}{remnant massive neutron star}
\acrodef{SASI}{standing accretion shock instability}
\acrodef{SGRB}{short $\gamma$-ray burst}
\acrodef{SPH}{smoothed particle hydrodynamics}
\acrodef{SN}{supernova}
\acrodefplural{SN}[SNe]{supernovae}
\acrodef{SNR}{signal-to-noise ratio}
\acrodef{ZAMS}{zero age main sequence}

\bibliographystyle{aasjournal}

\begin{appendix}
\section{GW probe of the one-armed spiral instability}\label{app:fig1_just}

\begin{figure*}[htb]
\centering
\includegraphics[width=0.49 \textwidth]{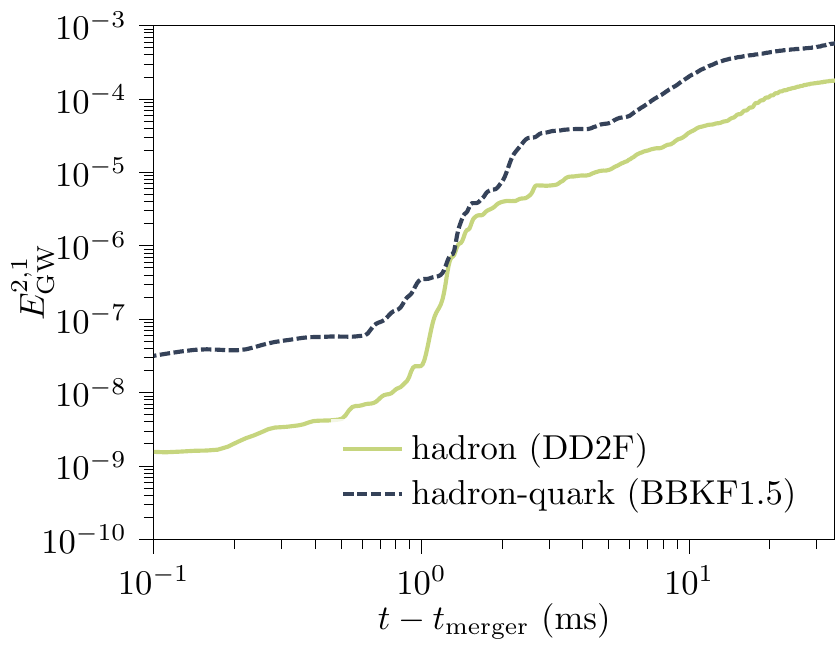}
\includegraphics[width=0.49 \textwidth]{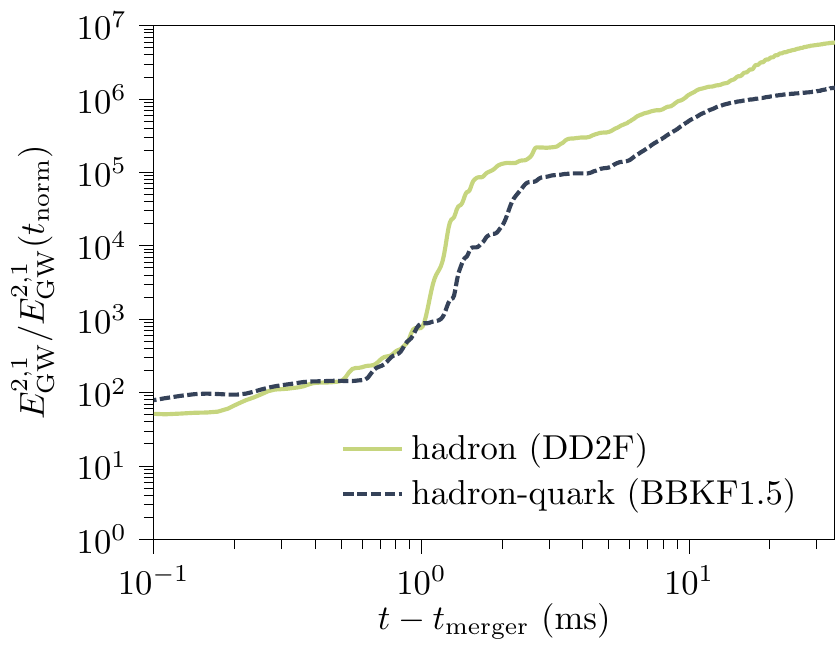}
\caption{\textit{Left panel}: Energy in the $l=2, m=1$ GW mode as a function 
of time for simulations employing a hadronic (DD2F) and hadron-quark 
(BBKF1.5) EOS; the simulations use identical initial conditions and are run 
with a grid resolution of $\Delta x = \SI{369.2}{m}$ in the finest grid. 
These results showcase that the one-armed spiral instability may be seeded at 
different levels in the postmerger environment for different simulations. 
\textit{Right panel:}
Same quantity as the left panel, but normalized to the value at a time 
shortly after merger, $t_{\rm norm} = t_{\rm merger} + \SI{0.5}{ms}$. 
Normalizing at this time accounts for the one-armed spiral instability being 
seeded at disparate levels across simulations.
}
\label{fig:fig1_just}
\end{figure*}

In Fig.~\ref{fig:energy_GW_egap} we depict an example of GW quantities that 
exhibit the suppression of the one-armed spiral instability for simulations that 
employ hadron-quark EOSs. In particular, we calculate the GW energy carried 
in the $l=2,m=1$ mode.
In Fig.~\ref{fig:energy_GW_egap} we show $E_{\rm GW}^{2,1}$ normalized to 
its value at a time shortly after the merger;
we depict normalized quantities because of the 
variable nature in which the one-armed spiral instability is seeded in the 
immediate post-merger environment in the context of numerical studies. 
Unless it is explicitly 
excited as a non-axisymmetric perturbation of a known amplitude (e.g., as a 
fixed-amplitude perturbation in the rest mass density), the 
one-armed spiral instability arises numerically from error at the 
level 
of floating-point precision~\citep{Espino:2019xcl}. 
As such, small differences in the early post-merger 
evolution of the fluid can result in the instability being seeded at 
different 
strengths. We do not explicitly seed the one-armed spiral instability using 
fluid 
perturbations in this work and, as a result, simulations that either run on 
different machines, use different grid resolutions, or use different 
numerical 
libraries result in different strengths for the initial instability seed. 
In Fig.~\ref{fig:fig1_just} we show the energy in the $l=2,m=1$ GW mode 
$E_{\rm GW}^{2,1}$ as a 
function of time for a set of low resolution simulations used to produce the 
error bars of Fig.~\ref{fig:energy_GW_egap}. 
The left panel of Fig.~\ref{fig:fig1_just} shows $E_{\rm GW}^{2,1}$ as 
extracted from our simulations and appears to show that the simulation 
employing a hadron-quark EOS produces a larger energy in the $l=2, m=1$ GW 
mode. However, it is clear the energy at a time shortly after the merger
$E_{\rm GW}^{2,1} (t_{\rm merger} + \epsilon)$ 
(where $\epsilon$ is a small additive time)
is larger for the hadron-quark simulation, suggesting that the one-armed 
spiral instability was seeded at a larger amplitude in that case. In order to account for the different levels at which the one-armed spiral 
instability is seeded in the immediate post-merger environment, we normalize 
the quantities depicted in Fig.~\ref{fig:energy_GW_egap} at a time 
shortly \emph{after} the 
merger $t_{\rm norm} = t_{\rm mer} + \epsilon$. 
We find that setting $\epsilon= \SI{0.5}{ms}$ 
results in all simulations in our work having roughly equal values of 
$E_{\rm GW}^{2,1}$ in the few ms immediately following merger. Normalizing 
at a time shortly after merger ensures that all simulations have approximate 
parity in the level at which the one-armed spiral instability is seeded and 
leads to the robust trend established in the right panel of 
Fig.~\ref{fig:energy_GW_egap}, regardless of grid resolution used.

\end{appendix}

\end{document}